\documentstyle[pre,aps,epsf]{revtex}
\newcommand{\bm}[1]{\mbox{\boldmath $#1$}}

\title{Dispersion of passive tracers in velocity field with non
delta-correlated noise} 
\author{P.~Castiglione and A.~Crisanti\\
\small Dipartimento di Fisica, Universit\`a di Roma ``La Sapienza'',
and INFM sezione Roma I \\ \small P.le A. Moro 2, 00185 Roma, Italy. }
\author{(Physical Review E, in press)}

\begin{document}
\maketitle \date{}

\begin{abstract}
The diffusive properties in velocity fields whose small scales are
parameterized by non $\delta$-correlated noise is investigated using
multiscale technique.  The analytical expression of the eddy
diffusivity tensor is found for a $2$D steady shear flow and it is an
increasing function of the characteristic noise decorrelation time
$\tau$.  In order to study a generic flow ${\bm v}$, a small-$\tau$
expansion is performed and the first correction ${\cal O}(\tau)$ to
the effective diffusion coefficients is evaluated.  This is done using
two different approaches and it results that at the order $\tau$ the
problem with a colored noise is equivalent to the white-in-time case
provided by a renormalization of the velocity field ${\bm v} \mapsto
\tilde{\bm v}$ depending on $\tau$.  Two examples of $2$D
closed-streamlines velocity field are considered and in both the cases
an enhancement of the diffusion is found.
\end{abstract}
PACS number(s): ; \\

\section{Introduction}

The problem of diffusion in a given velocity field has both
theoretical and practical relevance in many different fields of
science and engineering as, e.g., transport processes in chemical
engineering and combustion studies \cite{HKM}.  The tracers transport,
in particular the evolution of their concentration, plays an important
role in many aspects of geophysics.  For the oceanic flows, satellite
data indicate that the mesoscale features, like eddies and cold
filaments, advect temperature and nutrients over spatial and temporal
scales longer than those of the geostrophic turbulence.  The diffusion
enhancement by a given velocity field has attracted a lot of works in
the last years.  In particular the role of the velocity field
properties has been largely investigated while the effects of small
scales parameterization are not understood.
 
In this paper we will focus on the effects of a finite noise
correlation time. This problem is relevant in studying the transport
in the ocean since in this system the noise term comes from unresolved
velocity scales which are correlated in time.
   
In Section 2, by using the multiscale technique, we study the
diffusion properties of the model proposed in Ref. \cite{Lacpur} for
transport in the upper mesoscale ocean. The transport is described by
a Langevin equation with a Gaussian colored noise in time.
 
The aim is to understand whether a finite noise correlation time
$\tau$ enhances or depresses the dispersion process in a given
velocity field ${\bm v}({\bm x},t)$ with respect to the delta-correlated 
case ($\tau=0$).

Exploiting the scale separation in the dynamics we derive, using the
multiscale technique \cite{BLP78}, an effective diffusive equation for
the macrodynamics, the calculation of the effective diffusivity
second-order tensor is reduced to the solution of one auxiliary
partial differential equation
\cite{Piretal},\cite{Majetal},\cite{BCVV95}.
  
In Section 3 we consider a shear flow, in this case the diffusion
coefficient increases with $\tau$.  The solution of the auxiliary
equation is, in general, quite difficult, therefore, to investigate
the role of the finite $\tau$ in Section 4 we perform a small--$\tau$
expansion. An alternative method is presented in the Appendix A.
 
In Section 5 we study the case of two closed-streamlines fields that
mimics the transport in the Rayleigh-B\'enard system: the
quasi-two-dimensional flow studied by Shraiman in \cite{Shraiman} and
the AB flow.  In both the cases the presence of a small correlation
time enhances the diffusion process.

Conclusions are reserved for the final Section 6.

\section{Effective diffusion equation for Gaussian colored noise}

We consider large scale and long time dynamics of the model proposed
in \cite{Griffa} and already studied in \cite{Lacpur} for the
transport of a fluid particle in the upper mesoscale ocean:
\begin{equation}
{d \over dt} {\bm x} = {\bm v}({\bm x}, t ) + {\bm s }( t )
\label{Lan1}
\end{equation}
where ${\bm v}$ is a $d$-dimensional incompressible velocity field
(${\bm \nabla}\cdot {\bm v}=0$), for simplicity, periodic both in
space and in time and ${\bm s}$ is a Gaussian random variable of zero
mean and correlation function
\begin{equation}
\langle s_i ( t ) \; s_j (t') \rangle = \frac{\sigma^2}{\tau} \;
 \delta_{i j} e^{- \frac{ \mid t-t' \mid }{\tau}} .
\label{Lan11}
\end{equation}
The term ${\bm v}({\bm x},t)$ represents the part of the velocity
field that one is able to resolve, i.e., the larger scale mean flow,
whereas ${\bm s}(t)$ represents the part of the velocity field
containing the subgridscale flow processes, e.g. the small-scale
turbulence.  The plausibility of such a description is discussed in
\cite{Colin},\cite{Krauss},\cite{Zambia}.\\ In the limit $\tau \to 0$,
resulting $e^{-|t-t'|/\tau}/\tau \to \delta(t-t')$, the (\ref{Lan11})
reproduces the widely studied delta-correlated case
\begin{equation}
\langle s_i ( t ) \; s_j (t') \rangle =
\sigma^2\; \delta_{i j} \delta(t-t')
\label{Lan00}
\end{equation} 
the diffusive properties of which we would like to compare with the
$\tau$-correlated noise case.

To study the dispersion of tracers evolving according to
eqs. (\ref{Lan1}) and (\ref{Lan11}) on large scales and long times we
use the multiscale technique. This is a powerful mathematical method,
also known as homogenization, for studying transport processes on time
and spatial scales much larger than those of the velocity field ${\bm
v}$.  It has been already applied to the delta-correlated case
\cite{BLP78} and it has been shown that the motion on large time and
spatial scales is diffusive and it is described by an effective
diffusion tensor which takes into account the effects of the advecting 
velocity on the bare diffusion coefficient $\sigma^2$.

To apply this method to the case of Gaussian colored noise, we first
write eqs. (\ref{Lan1}) and (\ref{Lan11}) into a Markovian process by
enlarging the state space considering ${\bm s}(t)$ as a variable
evolving according to the Langevin equation:
\begin{equation}
{d \over dt} {\bm s} = - {1 \over \tau} {\bm s} + {\bm w }( t )
\label{Lan2}
\end{equation}
where now the noise ${\bm w }( t )$ is a white noise with correlation
functions
\begin{equation}
\langle w_i(t) \; w_j(t') \rangle \;=\; 2 \left( \frac{\sigma}{\tau}
\right)^2 \; \delta_{ij} \; \delta(t-t').
\label{correb}
\end{equation}
We have now a
two-variable (${\bm x},{\bm s}$) Markovian process whose 
associated Fokker-Planck equation can be easily obtained. Indeed introducing 
\begin{equation}
 {\bm Y} = \left( \begin{array}{c}
            {\bm x}\\{\bm s} 
       \end{array} \right) \; ; \;\;
 {\bm W} = \left( \begin{array}{c}
            {\bm w}\\{\bm w} 
       \end{array} \right)\; ; \;\;
 {\bm V} = \left( \begin{array}{c}
            {\bm v}+{\bm s}\\-{1 \over \tau}{\bm s} 
       \end{array} \right)\; ; \;\;
 {\widehat{A}}=\left( \begin{array}{cc}
            {\bm 0} &  {\bm 0} \\
            {\bm 0} &  {\bm 1} 
       \end{array} \right)
\end{equation}
the equations (\ref{Lan1}) and (\ref{Lan2}) 
become
\begin{equation}
{d \over dt} {\bm Y} = {\bm V} ({\bm Y}, t ) + {\widehat{A}} \cdot 
{\bm W }( t ).
\label{Lan3}
\end{equation}
The associated Fokker-Planck equation is 
\begin{equation}
\partial_t \Theta=\left( {\sigma \over \tau}\right)^2 {\bm \partial_{ss}}^2 
\Theta-({\bm v}+{\bm s}) \cdot {\bm \partial}_{x} \Theta +{1 \over \tau} 
\;{\bm \partial_s} \cdot ({\bm s}  \; \Theta)
\label{FP}
\end{equation}
where $\Theta=\Theta({\bm x},{\bm s},t)$ denotes the probability density.

The doubling of the space dimension is the price to pay for having a 
Fokker-Planck equation.
In the Appendix A we discuss a different approach to the problem which does 
not double the dimension of the space, but leads in general to a 
non-Markovian master equation.  

We can now apply the multiscale technique. Following \cite{BCVV95} 
in addition to the {\em fast} variables ${\bm x}$ and $t$ 
we introduce the {\em slow} variables defined as ${\bm X}=\epsilon {\bm x}$ 
and $T=\epsilon^2 t$ where $\epsilon \ll 1$ is the parameter controlling the
separation between the small scales related to the velocity field ${\bm v}$ 
and the large scale related to the $\Theta$ variation.
The two sets of variables are considered independent and so we have to make
the substitution
\begin{equation}
{\bm \partial_x} \mapsto {\bm \partial_x}+\epsilon 
{\bm \partial_X}
\;\;\; ; \;\;\; 
{\bm \partial}_t \mapsto {\bm \partial}_t+\epsilon^2 
{\bm \partial}_T.
\label{newgrad}
\end{equation}  
The solution of the Fokker-Planck equation (\ref{FP}) is sought as a 
perturbative series
\begin{equation}
\Theta({\bm x},t,{\bm X},T,{\bm s})=
\Theta^{(0)}+\epsilon \Theta^{(1)}+\epsilon^2 \Theta^{(2)}+...
\label{newteta}
\end{equation}
where the functions $\Theta^{(n)}$ depend on both fast and slow variables.
By inserting (\ref{newgrad}) and (\ref{newteta}) into the Fokker-Planck 
equation (\ref{FP}), 
equating terms of equal powers in $ \epsilon $ and choosing the solutions 
which have the same periodicities as the velocity field, we obtain a hierarchy 
of equations the first three of which are: 
\begin{eqnarray}
{\cal D} \; \Theta^{(0)} &=& 0 \label{eq0}\\
{\cal D} \;\Theta^{(1)}&=&-({\bm v}+{\bm s}) \cdot {\bm \partial_X} 
\Theta^{(0)} 
\label{eq1} \\
{\cal D} \;\Theta^{(2)}&=&-({\bm v}+{\bm s}) \cdot {\bm \partial_X} 
\Theta^{(1)} -\partial_T \Theta^{(0)}
\label{eq2}
\end{eqnarray}
where the operator ${\cal D}$ is defined as
\begin{equation} 
{\cal D} = \partial_t +({\bm v}+{\bm s}) \cdot {\bm \partial_x}
-{1 \over \tau} \;{\bm \partial_s}\;( {\bm s} \; \;)
-\left( {\sigma \over \tau} \right)^2 {\bm \partial_{ss}}^2. 
\label{opD}
\end{equation}
In order to solve eq. (\ref{eq0}) we make use of scale separation and 
we write the solution $\Theta^{(0)}$ as the sum of two terms: $\langle 
\Theta^{(0)} \rangle ( {\bm X},T,{\bm s} )$ depending on 
the {\em slow} variables  and  
 ${\widetilde{\Theta}}^{(0)} ( {\bm x}, t, {\bm s} )$
 depending on the {\em fast variables}. 
Here and in the following the  $\langle \;\cdot \;\rangle$
 indicates the average over the 
{\em fast} variables.  

Equation (\ref{eq0}) then splits into the two equations
\begin{equation}
{\cal D} \;{\widetilde{\Theta}}^{(0)} = 0
\label{eqtilde}
\end{equation}
\begin{equation}
\left( {\sigma \over \tau} \right)^2 {\bm \partial_{ss}}^2 
\langle \Theta^{(0)} \rangle
+{1 \over \tau} \;{\bm \partial_s}\;({\bm s} \; 
\langle \Theta^{(0)} \rangle) =0.
\label{eqmed}
\end{equation}
One can show \cite{BCVV95} that the solution ${\widetilde{\Theta}}^{(0)}$ will 
relax to a constant with respect to fast variables, so we can simply take 
\begin{equation}
\Theta^{(0)}({\bm x},t,{\bm X},T,{\bm s}) =
\langle \Theta^{(0)}\rangle({\bm X},T,{\bm s})
\end{equation} 
and write the solution of eq. (\ref{eqmed}) as
\begin{equation}
\langle \Theta^{(0)}\rangle = Q({\bm X},T) \; {\cal P}(s)
\label{solmed}
\end{equation}
where ${\cal P}(s)$ is defined as
\begin{equation}
{\cal P} (s) = \frac{ e^{-\tau s^2 / 2 \sigma^2}} 
{(2 \pi \sigma^2 / \tau)^{\frac{d}{2}}}
\label{P}
\end{equation}
with $d$ the dimension of the ${\bm x}$-space.
 
By using (\ref{solmed}) we see that eq. (\ref{eq1}) can be
written as 
\begin{equation}
{\cal D} \; \Theta^{(1)}= {\bm f}({\bm x},t,{\bm s}) \cdot {\bm G}({\bm X},T)
\end{equation}
with 
\begin{equation}
 {\bm f}({\bm x},t,{\bm s})=-({\bm v}+{\bm s})\;
{\cal P}(s) \; , \;\;\;\;\;\;\;\;
{\bm G}({\bm X},T)= {\bm \partial_X} Q({\bm X},T) 
\label{def}
\end{equation}           
with solution 
\begin{equation}
\Theta^{(1)}({\bm x},t,{\bm X},T,{\bm s})=
{\bm \chi}({\bm x},t,{\bm s}) \cdot {\bm G}({\bm X},T).
\label{sol1}
\end{equation}
The vector field ${\bm \chi}$ is called the auxiliary field and it solves 
the auxiliary equation
\begin{equation}
{\cal D} \;{\bm \chi}={\bm f}.
\label{eqaux}
\end{equation}
Finally by averaging eq. (\ref{eq2}) over the fast variables 
, $\langle \;\cdot \;\rangle$, and integrating over ${\bm s}$ 
, $\overline{(\;\cdot\;)}$,  we obtain  
\begin{equation}
{\bm \partial_X} \overline{\langle ({\bm v}+{\bm s}) \;\Theta^{(1)}\rangle}
=-\partial_T \overline{\langle \Theta^{(0)} \rangle}= - \partial_T 
Q({\bm X},T)
\label{eq2med}
\end{equation}
which, using eq. (\ref{sol1}), becomes
\begin{equation} 
\partial_T Q = -\overline{\langle (v_j +s_j) \; \chi_i \rangle}\; 
\partial_{X_i X_j}^2 Q = D^E_{ij}\;\partial_{X_i X_j}^2 Q.
\label{macrodi}   
\end{equation}  
This is the diffusion equation describing 
the large--scale dynamics, i.e., the dynamics in the slow variables. 
The effective eddy diffusivity tensor $D^E_{ij}$ 
is given by
\begin{equation}
D_{i j}^E = -\frac{1}{2}\; \left( \overline{ 
\langle (v_i + s_i) \; \chi_j \;\rangle} + \overline{ 
\langle (v_j + s_j) \; \chi_i \;\rangle}\right).
\label{diff}
\end{equation}
From the auxiliary equation (\ref{eqaux}) one can show 
that the $D^E_{ij}$ is positive definite.
Indeed, if we consider the $i$-th and the $j$-th component of 
(\ref{eqaux}), multiply by $ \chi_j $ and $ \chi_i $ 
respectively, sum the two terms and average the result over the 
periodicities, $\langle \;\cdot \;\rangle$, and
integrate over the random variable ${\bm s}$ we end up with
\begin{equation}
D_{ij}^E=\left( \frac{\sigma}{\tau} \right)^2 
\int d^d s \;\left( \partial_s \langle \chi_i \rangle \right)
\; \left( \partial_s \langle \chi_j \rangle \right) 
\;{\cal P}(s) \geq 0.
\label{defpos}
\end{equation} 
This result can be extended to non-periodic velocity field following 
the prescriptions in \cite{AvMa}.

\section{A solvable case: the stationary shear flow}

The resolution of the auxiliary field equation for a generic ${\bm v}$ 
is not an easy task. Therefore not trivial solvable cases are useful to 
understand the properties of the solution. In particular the auxiliary 
equation can be resolved for parallel flows, which in two dimensions have the form 
\begin{equation} 
{\bm v}(x,y)=(v(y), 0)
\label{shear}
\end{equation}
where $v(y)$ is an arbitrary function of $y$.
Note that these flows automatically satisfy ${\bm \nabla} \cdot {\bm v}=0$.
To evaluate the effective diffusion coefficients we first write the solution 
of th auxiliary equation as 
\begin{equation}
\chi_i({\bm x},t,{\bm s},{\bm x}',t',{\bm s}')=
-\int d t' \; d {\bm x}' \; d {\bm s}' \;{\cal G} 
({\bm x},t,{\bm s},{\bm x}',t',{\bm s}')\; 
\left(v_i({\bm x}')+s_i'\right) \; {\cal P} (s') 
\label{chigreen}
\end{equation}
where ${\cal G}$ is the Green function of the operator ${\cal D}$:
\begin{equation}
{\cal D}\;{\cal G}\;=\;\delta({\bm x}-{\bm x}') 
\;\delta(t-t') \;\delta({\bm s}-{\bm s}').
\label{eqgreen}
\end{equation}
Inserting (\ref{chigreen}) into eq. (\ref{diff}) we have:  
\begin{equation}
D_{i j}^E= - \int d t \; d {\bm x} \; d {\bm s} \; d t ' \; d {\bm x}' \; 
d {\bm s}' \;\left(v_i({\bm x})+s_i \right) \;{\cal G} 
({\bm x},t,{\bm s},{\bm x}',t',{\bm s}') 
\; \left(v_j({\bm x}')+s_j'\right)\;
 {\cal P}(s'). 
\label{diffgreen}
\end{equation}
Now we note that the Green function ${\cal G}$ can be written as
\begin{equation}
{\cal G}\;=\; \langle \delta({\bm x}-{\bm x}(t;{\bm x}',t')) 
\; \delta({\bm s}-{\bm s}(t;{\bm s}',t')) \rangle_{\bm w}
\label{green}
\end{equation}
where the average is over the realizations of the 
white noise ${\bm w}$, 
${\bm x}(t;{\bm x}',t')$ and ${\bm s}(t;{\bm s}',t')$ are the 
solutions of eqs. (\ref{Lan1}), (\ref{Lan2}) and (\ref{shear}) with initial 
condition ${\bm x}'={\bm x}(t';{\bm x}',t')$ and 
${\bm s}'={\bm s}(t';{\bm s}',t')$.
For the velocity field (\ref{shear}) the solutions of the equations 
(\ref{Lan1}) and (\ref{Lan2}) can be written as
\begin{equation}
\left\{
\begin{array}{l}
x(t) = x'+\int_{t'}^{t} d t_1 \; v(y(t_1)) +
\int_{t'}^{t} d t_1 \; s_1 (t_1)\\
y(t)= y'+\int_{t'}^{t} d t_1 \; s_2 (t_1)\\
{\bm s}(t) = {\bm s}' e^{-(t-t')/\tau}+\int_{t'}^{t} d t_1 \; {\bm w}(t_1) 
e ^{-(t-t_1)/\tau}.
\end{array}       
\right.
\label{solshe}
\end{equation}  
Inserting (\ref{green}) and (\ref{solshe}) into (\ref{diffgreen}), after 
some straightforward straightforward algebra one obtains:
\begin{equation}
D^{E}_{11}=\sigma^2+\frac{1}{2 \pi} \int d k \; \mid \widehat{v}(k)\mid^2 
\lim_{t \rightarrow \infty} \int_0^t dt'\; 
\exp 
\left(
- \sigma^2 \; k^2 \; \left[ 
(t-t') - \tau \left( 1-e^{ -\frac{t-t'}{\tau}} \right) 
\right] 
\right)
\label{diffshe}
\end{equation}
and
\begin{equation} 
D^{E}_{12}=0 \;\; , \;\;\; D^{E}_{22}=\sigma^2
\end{equation}
where $\widehat{v}(k)$ is the Fourier transform of $v(y)$.
The same result can be obtained directly from the 
definition 
\begin{equation}
D_{i j}^E = \lim_{t \rightarrow \infty } {1 \over 2 t} \langle \left( 
x_i (t)-\langle x_i \rangle \right) \; \left( x_j (t)- \langle x_j \rangle 
\right) \rangle
\label{cov}  
\end{equation} 
using (\ref{solshe}).  
Now, because of the inequality
\begin{equation} 
\exp \left[\sigma^2 \; k^2 \tau \left( 1-e^{-\frac{t-t'}{\tau}}\right)\right] 
\geq 1 
\label{disugua}
\end{equation}
one has 
\begin{eqnarray}
D_{11}^E (\tau) &\geq & \sigma^2 + \frac{1}{2 \pi} \int d k \; 
\mid \widehat{v}(k) \mid^2 
\lim_{t \rightarrow \infty} \int_0^t dt' \; e^{- \sigma^2 \; k^2 \; (t-t')} 
\nonumber \\
&=& \sigma^2 + \frac{1}{2 \pi} \int d k  
\frac{\mid \widehat{v}(k) \mid^2}{\sigma^2 \; k^2} = D_{11}^E (0) 
\label{zero}
\end{eqnarray}
Therefore for a stationary parallel flow a colored noise produces an 
enhancement of the dispersion.
Similar equations can be obtained for a time dependent shear flow. However 
in this case it is not simple to see the sign of the correction. 
The results will be reported elsewhere.
It is trivial to show that the results in this section hold also for a $3$-d 
shear flow :
\begin{equation}
{\bf v}(x,y,z)=\left( v(y,z), 0, 0 \right)
\label{shear3d}
\end{equation}   

\section{Eddy diffusivity for short noise correlation time}

By using multiscale technique, the calculation of the eddy diffusivities 
has been reduced to the solution of the auxiliary equation (\ref{eqaux}).
Numerical methods are generally needed to solve it  
but to do so we have to work in a $2 d$-dimensional space.
 In general, this is not feasible so, 
to get more insight the generic ${\bm v}$ case we study the small $\tau$ case
and expand the auxiliary field $\chi$ in a power series of 
$\tau$. 
A typical time of the physical system which can be compared to $\tau$ is 
$\tau_s = {\lambda}/{\langle v^2\rangle^{1/2}}$ i.e. the average time it 
takes a particle to travel a characteristic length 
$\lambda$.
 
Taking
\begin{eqnarray*}
\alpha_i &=& \sqrt{\tau}\; s_i \\
\chi_i &=& \frac{\tau}{2 \pi \sigma^2} \sum_{k=0}^{\infty} 
\chi_i^{\left( k \right)} ({\bm x},{\bm \alpha},t) \; \tau^{\frac{k}{2}}
\end{eqnarray*}
we obtain the following expression for the eddy diffusivity tensor
\begin{equation}
D_{i j}^E= -\frac{1}{2}\; \frac{1}{2 \pi \sigma^2} \left[
\sum_k \tau^{k \over 2} \left( \langle v_i \overline{\chi_j}^{(k)}
\rangle +\langle v_j \overline{\chi_i}^{(k)} \rangle \right)+
 {\tau}^{{k-1} \over 2} \left(\langle \overline{\alpha_i {\chi_j}^{(k)}}
\rangle +\langle \overline{\alpha_j {\chi_i}^{(k)}}\rangle \right) 
\right]
\label{difftau}
\end{equation}  
and for the auxiliary equation:
\begin{equation}
\sum_k 
\left[ \tau^{k \over 2} \left( \partial_t +{\bm v}\cdot {\bm \partial}_x
\right) 
+{\tau}^{{k - 1} \over 2} \; {\bm \alpha} \cdot {\bm \partial}_x
-{\tau}^{{k - 2} \over 2}\left( {\bm \partial_\alpha} ({\bm \alpha} \; \;) 
-\sigma^2 \partial^2_{\alpha \alpha}\right)
\right] {\chi_i}^{(k)}= - \left( v_i +\frac{\alpha_i}{\sqrt{\tau}} \right) 
e^{-\frac{\alpha^2}{2 \sigma^2}}. 
\label{auxtau}
\end{equation}  
From the expression (\ref{difftau}) we see that in order to determine the 
correction of order $\tau$ to $D_{i j}^E$,  we need the quantities 
$\langle {\chi_i}^{(0)} \rangle$, $\overline{\chi_i}^{(0)}$,  
$\langle {\chi_i}^{(1)} \rangle$, $\overline{\chi_i}^{(1)}$,
$\langle {\chi_i}^{(2)} \rangle$, $\overline{\chi_i}^{(2)}$,
$\langle {\chi_i}^{(3)} \rangle$.
The fields ${\chi_i}^{(k)}$ obey to the equations
\begin{eqnarray}
{\cal O}_\alpha {\chi_i}^{(0)} &=& 0 \label{opchi0} \\
{\cal O}_\alpha {\chi_i}^{(1)} &=& {\bm \alpha}\cdot {\bm \partial_x}
{\chi_i}^{(0)} + \alpha_i e^{-\frac{\alpha^2}{2 \sigma^2}}\label{opchi1}\\
{\cal O}_\alpha {\chi_i}^{(2)} &=& \left( \partial_t +{\bm v}\cdot 
{\bm \partial}_x \right) {\chi_i}^{(0)} +{\bm \alpha} \cdot {\bm \partial}_x 
{\chi_i}^{(1)} + 
v_i  e^{-\frac{\alpha^2}{2 \sigma^2}}  \label{opchi2} \\
{\cal O}_\alpha {\chi_i}^{(h)} &=& \left( \partial_t +{\bm v}\cdot 
{\bm \partial_x}\right) {\chi_i}^{(h-2)} +{\bm \alpha} \cdot {\bm \partial_x} 
{\chi_i}^{(h-1)} \;\;\;\;\;\;\;\; h \geq 3 \label{opchi3}
\end{eqnarray}
where the operator ${\cal O}_\alpha$ is defined as 
\begin{equation}
{\cal O}_\alpha = {\bm \partial_\alpha }\cdot({\bm \alpha} \; \;) - 
\sigma^2 \partial_{\alpha \alpha}^2.
\end{equation}
The solutions of eqs. (\ref{opchi0}), (\ref{opchi1}) and (\ref{opchi2}) 
can be written in the form
\begin{eqnarray}
{\chi_i}^{(0)} &=& \widetilde{\chi_i}^{(0)} ({\bm x},t)\; 
e^{-\frac{\alpha^2}{2 \sigma^2}} \label{solchi0}\\
{\chi_i}^{(1)} &=& \left( a_{1i} \alpha_1 + a_{2i} \alpha_2+a_{3i} \right) 
\; e^{-\frac{\alpha^2}{2 \sigma^2}} \label{solchi1}\\
{\chi_i}^{(2)} &=& \left( b_{1i} \alpha_1^2+ b_{2i} \alpha_2^2+
b_{3i} \alpha_1 \alpha_2+b_{4i} \alpha_1+b_{5i} \alpha_2+b_{6i} \right) \; 
e^{-\frac{\alpha^2}{2 \sigma^2}} \label{solchi2}    
\end{eqnarray}
the coefficients $a_{ij}$ and $b_{ij}$ are functions of 
${\bm x}$ and $t$ while for the field ${\chi_i}^{(3)}$ it results 
that 
\begin{equation} 
\langle {\chi_i}^{(3)} \rangle = c({\bm x},t) \;e^{-\frac{\alpha^2}{2 
\sigma^2}}. 
\label{chi3} 
\end{equation}
By inserting expressions (\ref{solchi0}), (\ref{solchi1}) and (\ref{solchi2}) 
into eqs. (\ref{opchi0}), (\ref{opchi1}), (\ref{opchi2}) and (\ref{opchi3}) 
we can determine the coefficients $a_{ij}$ for $i=1,2$ and $b_{ij}$ 
for $i=1,..,5$ 
and then by integrating over $\alpha$  the equations for 
${\chi_j}^{(2)}$, ${\chi_j}^{(3)}$ and ${\chi_j}^{(4)}$ respectively
we finally have the equations for the remaining functions 
${\widetilde{\chi}_i}^{(0)}$, $a_{3i}$ and $b_{6i}$ 
\begin{eqnarray}
{\cal O}_{xt} \; {\widetilde{\chi_i}}^{(0)} &=& - v_i  \label{eqcost1}\\ 
{\cal O}_{xt} \; a_{3i} &=& 0   \\
{\cal O}_{xt} \; b_{6i} &=& \frac{3}{2} \sigma^2 \left[ \partial^2 v_i + 
\partial^2 v_{j} \partial_{j}  {\widetilde{\chi}_i}^{(0)}\right]  
+ 2 \; \sigma^2 \partial_{j} v_{m} \; \partial_{j m}^2 
{\widetilde{\chi}_i}^{(0)}    
\label{eqcost2}
\end{eqnarray}
where
\begin{equation} 
{\cal O}_{xt} = \partial_t + {\bm v} \cdot {\bm \partial} -
\sigma^2 \partial^2.
\end{equation}

The $D^E_{ij}$ coefficients at the first order in $\tau$ read
\begin{eqnarray}
D^E_{ij} &=& \sigma^2 \delta_{ij} - \frac{1}{2} \left( \langle v_i\; 
{\widetilde{\chi}}_j^{( 0 )} \rangle +\langle v_j\; 
{\widetilde{\chi}}_i^{( 0 )} \rangle  \right) \nonumber \\
       &-& \frac{\tau}{2} \; \left[ \frac{\sigma^2}{2} \left( \langle v_i 
\; \partial^2 
{\widetilde{\chi}}_j^{(0)} \rangle +  \langle v_j \; \partial^2 
{\widetilde{\chi}}_i^{(0)} \rangle \right) + 
\langle v_i\; b_{6j} \rangle +\langle v_j\; b_{6i} \rangle  \right].
\label{diffprimo}
\end{eqnarray}
We note that instead of the $2d$-dimensional equation (\ref{diff}) 
we have now a system of two $d$-dimensional 
equations (\ref{eqcost1}) and (\ref{eqcost2})
without the random variable ${\bm s}$. Of course this is 
numerically much more convenient.

We note that by defining new velocity and auxiliary fields as
\begin{eqnarray}
\widetilde{\bm v}&=&{\bm v}-\frac{\sigma^2 \tau}{2} \partial^2 
{\bm v} \label{vtilde}\\  
\widetilde{\bm \chi}&=&\widetilde{\bm \chi}^{(0)}+\tau \left( 
{\bm b}_{6}+ \sigma^2 \partial^2 \widetilde{\bm \chi}^{(0)} \right)
\label{chitilde}
\end{eqnarray}
and negletting terms of ${\cal O}(\tau^2)$ equations (\ref{eqcost1}),
(\ref{eqcost2}) and (\ref{diffprimo}) can be written as 
\begin{equation}
\left( \partial_t + \widetilde{\bm v}{\bm \partial} -\sigma^2 
\partial^2\right)\; \widetilde{\chi}_i = \widetilde{v}_i
\label{auxtilde}
\end{equation}
and
\begin{equation}
D^E_{ij}=\sigma^2 \delta_{ij}- \frac{1}{2} \left( \langle \widetilde{v}_i 
\widetilde{\chi}_j \rangle + \langle\widetilde{v}_j \widetilde{\chi}_i \rangle \right)
\label{difftilde}
\end{equation}
formally equivalent to the Gaussian white noise result. 

In the Appendix A we use a different method to obtain the 
expression of the eddy diffusivity tensor for small $\tau$ up to 
${\cal O}(\tau^2)$.
By starting from the Master equation associated with the Langevin equation 
(\ref{Lan1}) and using a small $\tau$ expansion  
we derive the Fokker-Planck equation   
 \begin{equation}
\partial_t \Theta ({\bm x},t) = -\partial_i \left[ v_i ({\bm x},t) \; 
\Theta ({\bm x},t) \right] +
\; \partial_{ij}^2 \left[ {\cal D}_{ij} ({\bm x},t) \; \Theta ({\bm x},t)
\right]
\label{FPapp}
\end{equation}
with
\begin{equation}
{\cal D}_{ij}({\bm x},t) =\sigma^2\; \left[ \delta_{ij} + \frac{\tau}{2} 
\left[ \partial_i v_j ({\bm x},t)\;+ \;\partial_j v_i ({\bm x},t) \right]
\right].
\label{Dapp}
\end{equation}
Now, by applying multiscale technique to (\ref{FPapp}) we end up with the 
following equations

\begin{eqnarray}
{\cal O}_{xt} w_i^{(0)}&=&-v_i \label{chi0app}\\
{\cal O}_{xt} w_i^{(1)}&=&\sigma^2 \left( \;\partial^2 v_i + 
\;\partial^2 v_j \; \partial_j w_i^{(0)} +
\;\partial_k v_j \; \partial^2_{kj} w_i^{(0)} \right) 
\label{chi1app}\\
D^E_{ij} &=& \sigma^2 \delta_{ij} - \frac{1}{2} \left( \langle v_i\; 
w_j^{(0)} \rangle + \langle v_j\; w_i^{(0)} \rangle  \right) \nonumber \\
       &-& \frac{\tau}{2} \; \left[ \left(   
\langle v_i \; \partial^2 w_j^{(0)} \rangle  
+ \langle v_j \; \partial^2 w_i^{(0)} \rangle \right) +
\langle v_i \; w_j^{(1)} \rangle + \langle v_j \; w_i^{(1)} \rangle \right]
\label{diffapp}
\end{eqnarray} 
which differ from the previous ones.

This result poses the question about the validity of the two sets of 
equations and hence of the expansions. 
In order to answer to this question we firstly note that the two sets 
can be considered as a particular case of the {\em generalized} equations
\begin{eqnarray}
{\cal O}_{xt} w_i^{(0)}&=&-v_i \label{chi0g}\\
{\cal O}_{xt} w_i^{(1)}&=&\sigma^2 \left( (p-q) \;\partial^2 v_i + 
(p-q) \;\partial^2 v_j \; \partial_j w_i^{(0)} +
p \;\partial_k v_j \; \partial^2_{kj}w_i^{(0)} \right) 
\label{chi1g}\\
D^E_{ij} &=& \sigma^2 \delta_{ij} - \frac{1}{2} \left( \langle v_i\; 
w_j^{(0)} \rangle +\langle v_j\; 
w_i^{(0)} \rangle  \right) \nonumber\\
       &-& \frac{\tau}{2} \; \left[ 
q \sigma^2 \left( \langle v_i \; \partial^2 w_j^{(0)} \rangle +  
\langle v_j \; \partial^2 w_i^{(0)} \rangle \right) 
+ 
\langle v_i\; w_j^{(1)} \rangle +\langle v_j\; w_i^{(1)} \rangle  \right]
\label{diffgen}
\end{eqnarray} 
obtained starting from the Master equation 
\begin{equation}
\partial_t \Theta({\bm x},t)=-\widetilde{v}_i({\bm x},t)\; \partial_i 
\Theta({\bm x},t) + \partial_{ij}^2 \left[ {\cal D}_{ij}({\bm x},t) 
\Theta({\bm x},t) \right]  
\label{FPG}
\end{equation}
where
\begin{eqnarray}
\widetilde{v}_i &=& v_i +q\; \sigma^2 \tau \partial^2 \; v_i 
\label{vgtilde} \\
{\cal D }_{ij} &=& \sigma^2 \left( \delta_{ij} + 
\frac{p \tau}{2} \left( \partial_i \;v_j \;+ \;\partial_j \;v_i  
\right) \right)
\label{diffg}
\end{eqnarray}  
and by applying the multiscale technique. 

The role of the two free parameters $p$ and $q$ is of generalizing the
Master equation (\ref{FPapp}) being our feeling that there exists a
family of different microscopic process specified by $\tilde{\bf v}$ and 
${\cal D}_{ij}$ that correspond to the same macroscopic diffusive process
specified by $D^E_{ij}$.

The first set consisting of the equations (\ref{eqcost1}), 
(\ref{eqcost2}) and (\ref{diffprimo}) corresponds to $q=1/2$ and  
$p=2$, whereas the second one consisting of the equations 
(\ref{chi0app}), (\ref{chi1app}) and (\ref{diffapp}) corresponds 
to $q=0$ and $p=1$.
Now a closer analysis of the diffusion coefficients (\ref{diffg}) reveals 
that it may take negative values, introducing unphysical singularities into 
the problem. Moreover the multiscale technique requires a $D^E_{ij}$ definite 
positive.
This means that (\ref{FPG}) do not constitute genuine Markovian process 
with a well defined corresponding Langevin equation driven by a Gaussian 
white noise.
In general this expansion do not converge uniformly in ${\bm x}$ and the 
range of validity is restricted to $\tau \ll 1 $, $\tau/\sigma^2 \ll 1$ 
(in dimensionless unit), thus they are asymptotic estimate for 
$\tau \rightarrow 0$.
In this range of equations validity we do not have any 
{\em a priori} arguments for  choosing one of the two sets of equations 
and it is for this reason that we expect to obtain the same value 
 of the eddy diffusivity tensor using the two sets.  
This is indeed the case for the parallel shear flows (see Appendix B).  
This result can be thought as a first indication of the equivalence 
(with respect to the value of the diffusivity) of the 
{\em generalized} equations. 
 
In general we do not expect that all possible choices of $q$ and $p$ give 
the same $D^E_{ij}$ but we can think to select the class of equivalent 
process by applying the {\em generalized} equations to the parallel shear 
flow and imposing $D^E_{ij}$ equal to the known expression (\ref{zero}) 
independent on both $q$ and $p$. 
This calculation is reported in Appendix B and it ends up in the following 
condition 
\begin{equation}
p = 2 \; q + 1
\label{condshe}
\end{equation}

All these considerations suggest that for short $\tau$ there exists
a class of equivalent equations depending on the parameter $p$ 
that lead to the same eddy diffusivity tensor.

If this is the case, among the all possible choices of $q$ and $p$  
consistent with (\ref{condshe}), we can choose: $p=0$ and $q=-1/2$.
In this case ${\cal D}_{ij}=\sigma^2 \delta_{ij}$ and (\ref{FPapp}) 
reduces to a truly Markovian process with the associated Langevin equation
\begin{equation}
{d \over dt} {\bm x} = \widetilde{{\bm v}}({\bm x} , t ) + {\bm \xi }( t )
\label{Lan32}
\end{equation}
where ${\bm \xi}$ is a Gaussian white noise. 
This is the microscopic Markovian process which approximates the long time 
and large space transport properties of a colored noise process with short 
noise correlation time. 
In other words to study the diffusion properties of 
(\ref{Lan1})--(\ref{Lan11}) for small 
$\tau$ we can replace the original colored noise process with the process 
described by (\ref{Lan32}). 
This will give the correct diffusion coefficients up to ${\cal O}(\tau^2)$. 

We have checked numerically that for the AB flow the eddy diffusivity tensor 
assumes the same value for the three different choices of the parameters 
$q$ and $p$ consistent with (\ref{condshe}): 
$p=2$ and $q=1/2$, $p=1$ and $q=0$, $p=0$ and $q=-1/2$.
This make us confident with our conclusions.
 
\section{Flows with closed streamlines}
We apply now our analysis to two models for the 
Rayleigh-B\'enard steady convection: 
the first one consists of an horizontal extent of convection cells much 
larger than its height so that the flow can be considered 
quasi-two-dimensional; 
the second one is the two-dimensional AB flow made of a structure 
periodically repeated in the space.

\subsection{A quasi-two-dimensional flow}
We consider the flow discussed by Shraiman \cite{Shraiman}.
This is described by the stream function  

\begin{equation}
\psi(x,y)=\frac{v L}{\pi k} \sin \left( \frac{\pi k}{L} x
\right) \sin \left(  \frac{\pi}{L} y \right)
\label{stream}
\end{equation}
with $v$ being the characteristic velocity, $L$ the height of the cell
 ($y \in [0,L]$) and $L/k$ the $x$-periodicity of the roll pattern.
The top and the bottom plates of the cell are assumed impermeable for 
the passive scalar so that the appropriate boundary conditions for the 
tracers density function $\Theta$ are $\partial_y \Theta \mid_{y=0,L} =0$.
The streamlines of the flow are illustrated in Fig. (\ref{fig_RB}).
Using Fokker-Planck equation

\begin{equation}
\partial_t \Theta = -{\bm v}\cdot {\bm \partial} \Theta + \sigma^2 
\partial^2 \Theta,
\label{FPO}
\end{equation}
for large P\'eclet number
\footnote{The case studied by Shraiman corresponds to the
delta-correlated case ( $\tau=0$ )
$$\langle s_i ( t ) \; s_j (t') \rangle = \sigma^2\; \delta_{i j}
\delta(t-t') $$ where $\sigma^2$ is now the molecular diffusivity of
the considered system.} $Pe=v L/\sigma^2$, $k=2$ and $\tau=0$ the eddy
diffusivity coefficient $D^E_{11}(0)$ has been calculated in
\cite{Shraiman} and it is
\begin{equation}
D^E_{11}(0)=\frac{\sigma^2}{\sqrt{\pi}} \; \sqrt{Pe}= 
\sqrt{ \frac{v L \sigma^2}{\pi} }.
\label{shradiff}
\end{equation}
According to the results of the previous section, the colored noise case with 
small $\tau$ is described up to order ${\cal O}(\tau^2)$ by the same
Fokker-Planck equation provided the velocity field is renormalized as eq. 
(\ref{vgtilde}) with $q=-1/2$.
Taking into account that 
$v_x=-\partial_y \psi$, $v_y=\partial_x \psi$, 
and using (\ref{stream}), we have       
$$\partial^2 \psi = -(\pi /L)^2 \;(1+k^2) \;\psi.$$ 
Therefore up to order ${\cal O}(\tau^2)$ the diffusion is described by 
\begin{equation}
\widetilde{\psi}(x,y)=\left( 1 + \frac{ \sigma^2 \tau}{2} 
\left( \frac{\pi}{L} \right)^2 \left( 1 + k^2 \right) \right) 
\psi(x,y) = c \; ( \sigma^2,\tau,k ) \; \psi(x,y). 
\label{streamnew} 
\end{equation}
Since $c$ does not depend on ${\bm x}$ and $t$, we can repeat the calculation 
of Shraiman and obtain under the same conditions of (\ref{shradiff}) the 
expression for the eddy diffusivity coefficient    
\begin{eqnarray}
D^E_{11}(\tau)&=&
\frac{\sigma^2}{\sqrt{\pi}}\; \sqrt{ c \; Pe}= 
D^E_{11}(0)\; \sqrt{ \left( 1 + \frac{5 \;\sigma^2 \tau}{2} 
\left( \frac{\pi}{L} \right)^2  \right) }\nonumber \\
&=&D^E_{11}(0)\;\left( 1 + \frac{5 \;\sigma^2 \tau}{4} 
\left( \frac{\pi}{L} \right)^2  \right) +{\cal O}(\tau^2).
\label{shradifftau}
\end{eqnarray}
We then conclude that, in this case, a {\em small} $\tau$ enhances the 
diffusion coefficient.

The same result can be deduced from the multiscale equations
(\ref{auxtilde}) and (\ref{difftilde}): in fact because of the
structure of these two equations it is not difficult to show that if
we change only the module of the velocity field ($\widetilde{\bm v}=c
{\bm v}$) and we know the explicit form of $D^E_{ij}=f({\bm v})$ as a
function of ${\bm v}$ we have $\widetilde{D}^E_{ij}=f(\widetilde{\bm
v})$. For large P\'eclet number and $\tau=0$ the function $f({\bm v})$
is given by eq. (\ref{shradiff}) from which (\ref{shradifftau})
follows.

\subsection{The AB flow}
The AB flow is given by the velocity field 
\begin{equation}
v(x,y)=(B \cos(y),\;A \cos(x)).
\label{BC}
\end{equation}
For $A=B=1$ the streamlines form a closed periodically repeated
structure made of four cells as shown in Fig. (\ref{fig_AB}).  We
expect that the diffusive behavior of such a system is similar to the
previous case, in fact we know \cite{SoloGollu} that for small
P\'eclet number $Pe$ the eddy diffusivity tensor is proportional to
$\sqrt{Pe}$ like in the quasi-two-dimensional case.

\begin{figure}[ht]
\epsfxsize=220pt\epsfysize=183.68pt\epsfbox{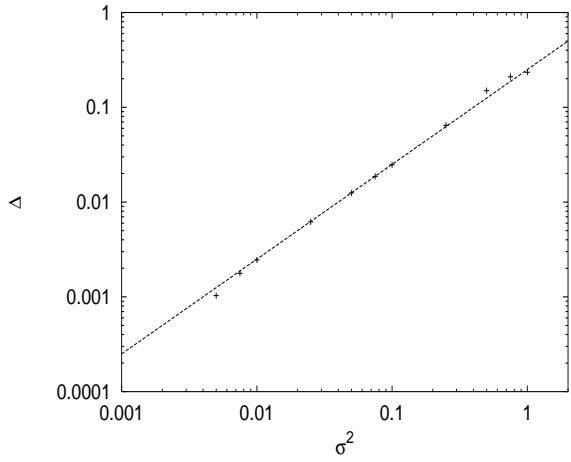}
\caption{ The ratio $\Delta$ as a function of the $\sigma^2$ for the
two-dimensional $AB$ flow with $A=B=1$.  The continuous line is the
prediction obtained by (\protect\ref{diffstar}) while the points are
the numerical results obtained from the (\protect\ref{diffnum}). All
the quantities here are supposed to be dimensionless.  }
\label{fig1}
\end{figure}

\begin{figure}[ht]
\epsfxsize=220pt\epsfysize=183.68pt\epsfbox{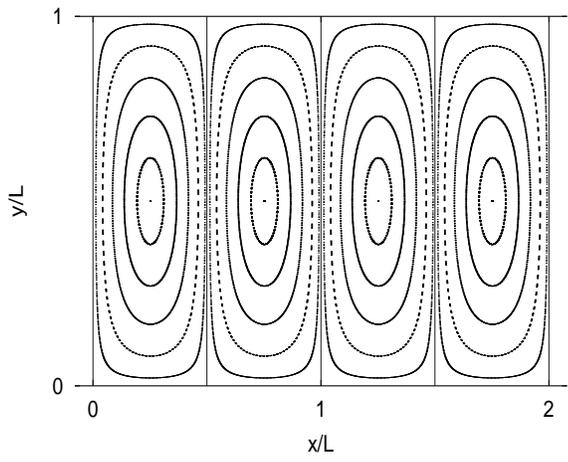}
\caption{ The streamlines for the {\em quasi} two-dimensional 
flow (\ref{stream}) with $k=2$.}
\label{fig_RB}
\end{figure}

\begin{figure}[ht]
\epsfxsize=220pt\epsfysize=183.68pt\epsfbox{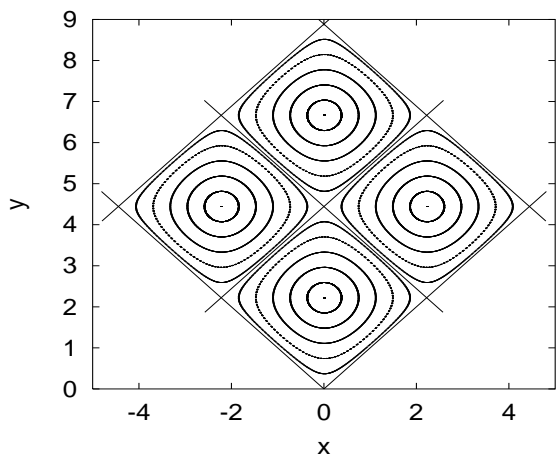}
\caption{ The streamlines for the two-dimensional $AB$ flow (\ref{BC})
with $A=B=1$. $x$, $y$, $A$ and $B$ are supposed to be all
dimensionless.  }
\label{fig_AB}
\end{figure}

In the Figure (\ref{fig1}) the behavior of the quantity $\Delta=
[ D^E(\tau)-D^E(0) ] / [ D^E(0) \; \tau ]$ versus $\sigma^2 $ is
shown. The correction $\Delta$ has been calculated by integrating
numerically the equations (\ref{eqcost1}) and (\ref{eqcost2}) and
evaluating the quantity

\begin{equation} 
- \frac{1}{D^E(0)} \; \left[ \frac{\sigma^2}{2}\; 
\langle v_1 \partial^2 {\widetilde{\chi}}_1^{(0)} \rangle + 
\langle v_1\; b_{61} \rangle \right]\; =
\; \frac{D^E(\tau)-D^E(0)}{D^E(0) \; \tau}=\Delta 
\label{diffnum}
\end{equation}
for different values of $\sigma^2$.
The equations are solved by using a pseudo-spectral method \cite{Gottl77} 
in the basic periodicity cell with a grid mesh of $64 \times 64$ points. 
De-aliasing has been obtained by a proper circular truncation which 
ensures better isotropy of numerical treatment.

It is evident that also in this case the numerical results follow a
linear behavior with a positive angular coefficient and so we can
conclude that the introduction of the colored noise leads to an
enhancement of the diffusion.  In particular we can see that the
numerical results follow very well the line $\Delta=\sigma^2 /
4$. This is not surprising because we know that for large $Pe$
\begin{equation}
D^E(0)=D_{11}^E(0)=D_{22}^E(0)= C_1\; \sqrt{Pe}=C_2 \; \sqrt{v}
\label{tau0}
\end{equation}
therefore using the same arguments of the previous section we can 
deduce that 
\begin{equation}
D^E(\tau)\;=\;
D^E(0)\; \sqrt{ 1 + \frac{1}{2} \sigma^2 \; \tau} \;= \; 
D^E(0)\; \left( 1+\frac{1}{4} \sigma^2 \tau + {\cal O} (\tau^2) \right).
\label{diffstar}
\end{equation}
in a very good agreement with the numerical results.

\section{Summary and Conclusions}

In this paper we have studied the transport properties in velocity
fields whose small scales are parameterized by Gaussian colored noise.
We analyzed in particular the effects of a finite noise correlation
time $\tau$ on the diffusive properties for large time and spatial
scales. In this limit, using the multiscale technique, we derive the
diffusion equation (\ref{macrodi}) and the associated effective
diffusion tensor.  The latter is obtained, once the velocity field
${\bm v}$ is given, by the solution of an auxiliary equation, see
(\ref{eqaux}) and (\ref{diff}), of the same structure of the original
Fokker-Planck equation \cite{Risken}.  The former is, however, an
exact result for the diffusive regime valid for very long times, thus
avoiding all finite time effects of the Fokker-Planck equation or the
associated Langevin equation (\ref{Lan1}),(\ref{Lan11}).

The auxiliary equation cannot be solved for a generic velocity field,
nevertheless there are nontrivial flows for which the solution can be
found. This is the case for the steady parallel flow for which the
effective diffusion coefficient is an increasing function of $\tau$.
To study in more details the effects of a small noise correlation time
for a generic ${\bm v}$ we have performed a small-$\tau$ expansion and
evaluated the first correction ${\cal O}(\tau)$ to the effective
diffusion coefficient.  This is done by using two different
approach. We find that to order ${\cal O}(\tau)$ there exist a
one-parameter family of flows with the same diffusion properties.
This invariance can be used to pick up the most convenient microscopic
dynamics, from both analytical and numerical porpoises.  We apply the
small-$\tau$ results to two two--dimensional model flows with closed
streamlines. In both the cases we find an enhancement of the
diffusion.

The enhancement of the diffusion for small $\tau$ has been interpreted
in \cite{AnPa} in terms of interference mechanism between turbulent
and molecular diffusion.  The colored noise makes the diffusion
particles forgotten of their previous positions less rapidly than in
the white noise case, thus the Lagrangian correlation time increases
and so does the eddy-diffusivity.

The study of the problem for $\tau$ not small is object of current 
work.

 
\section*{Acknowledgments}
We warmly thank Angelo Vulpiani, Massimo Vergassola and Andrea Mazzino
for extensive, stimulating and very useful discussions and
suggestions.  PC is grateful to the European Science Foundation for a
TAO exchange grant.  This work has been partially supported by INFM
(Progetto Ricerca Avanzata- TURBO) and by MURST (program
no. 9702265437).
   
\appendix
\section{The master equation for colored noise and small $\tau$}

In this appendix we derive the master equation for the probability
density $\Theta({\bm x},t)$ in the limit of small $\tau$ for the
process described by the Langevin equation (\ref{Lan1}) and
(\ref{Lan11}).  If the random variable ${\bm s}$ is a Gaussian white
noise of zero mean, $\Theta({\bm x},t)$ satisfies the Fokker-Planck
equation. We address here the case of a colored noise. Now the process
${\bm x}$ is non-Markovian and no exact simple equation for $\Theta $
is known.  Let us consider an Ornstein-Uhlenbeck process ${\bf s}$,
that is a zero mean Gaussian process with correlations [c.f.r
eq. (\ref{Lan11})]
\begin{equation}
C_{ij}(t,t')=\langle s_i(t) \;s_j(t')\rangle=
\frac{\sigma^2}{\tau} \; \delta_{i j} e^{- \frac{ \mid t-t' \mid }{\tau}} 
\label{corre}
\end{equation}
where $\tau$ is the correlation time.
The probability density is given by 
\begin{equation}
\Theta({\bm x},t)=\langle \delta ({\bm x}(t) -{\bm x}) \rangle
\label{densi}
\end{equation}
where ${\bm x}(t)$ is a solution of (\ref{Lan1}) for a given
realization of ${\bm s}$ and for a given initial condition. The
average is taken over the noise realizations.  Taking the time
derivative of (\ref{densi}) and using eq. (\ref{Lan1}) one gets
\begin{equation}
\partial_t \Theta({\bm x},t)= -\partial_i \left[ v_i ({\bm x},t) 
\Theta({\bm x},t) \right] + \partial_i \langle s_i(t) \delta({\bm x}(t)-
{\bm x})\rangle.
\label{dedensi}
\end{equation}
Taking advantage of the Gaussian nature of ${\bm s}$ the average in
(\ref{dedensi}) can be rewritten as
\begin{equation}
\partial_t \Theta({\bm x},t)= -\partial_i \left[ v_i ({\bm x},t) \; 
\Theta ({\bm x},t) \right] +
\; \partial_{ij}^2 \;\int_{t_0}^t dt' \; \left( \frac{\sigma^2}{\tau}\right)
e^{-\frac{t-t'}{\tau}}
\left< \frac{\delta x_j(t)} {\delta s_i(t')} \; \delta({\bm x}(t)-{\bm x}) 
\right>.
\label{FPC}
\end{equation} 
Because of the $\delta$-function a closed equation is possible only if
the functional derivative either does not involve the process ${\bm
x}$ or depends on it solely at the ``Markovian'' end-point $t=t'$.  At
this stage thus we cannot simplify the master equation any further.
In the limit of small correlation time $\tau$ a closed equation can be
derived by performing the largely used small $\tau$ expansion.  If the
noise is close to the white noise limit ($\tau=0$) it is reasonable to
expand the functional derivative about its Markovian value, i.e., the
one obtained for the $\delta$-correlated noise.  The Taylor expansion
of $\delta x_j(t) / \delta s_i(t')$ around the Markovian end point
$t'=t$ is
\begin{eqnarray}
\frac{\delta x_j(t)} {\delta s_i(t')}&=&
 \left. {\delta x_j(t) \over \delta s_i(t')} \right|_{t'=t}  \;+\;
\left. {d \over d t'} {\delta x_j(t) \over \delta s_i(t')}  \right|_{t'=t}  
\;(t'-t)\; +\; ... \nonumber\\
&=& \delta_{ij}-\partial_i v_j ({\bm x},t) \; (t'-t) \;+\cdots
\label{FPexpa}
\end{eqnarray}
Inserting the expansion (\ref{FPexpa}) into (\ref{FPC}), keeping only
the first terms in $\tau$ and negletting the transients, i.e. the
letting to $-\infty$ we obtain after a straightforward algebra the
small $\tau$ master equation:
\begin{equation}
\partial_t \Theta ({\bm x},t) = -\partial_i \left[ v_i ({\bm x},t) \; 
\Theta ({\bm x},t) \right] +
\; \partial_{ij}^2 \left[ {\cal D}_{ij} ({\bm x},t) \; \Theta ({\bm x},t)
\right]
\label{FPfinale}
\end{equation}
where
\begin{equation}
{\cal D}_{ij}({\bm x},t) =\sigma^2\; \left[ \delta_{ij} + \frac{\tau}{2} 
\left( \partial_i v_j ({\bm x},t)\;+ \;\partial_j v_i ({\bm x},t) \right)
\right]
\label{D}
\end{equation}
and use of incompressibility has been made.

We note that this expansion does not converge uniformly in ${\bm x}$,
and the diffusion coefficient (\ref{D}) may exhibit negative values,
thereby introducing unphysical singularities into the problem. In
other words (\ref{FPfinale}) and (\ref{D}) does not constitute a truly
Markovian process with well-defined corresponding Langevin equation
driven by white noise.  In general these equations are valid only for
$\tau \ll 1$ and $\tau / \sigma^2 \ll 1$ (in dimensionless units).

\section{The generalized formulas in the steady shear flow case}

The exact expression of the eddy diffusivity tensor for a stationary
bidimensional shear flow
\begin{equation}
{\bm v}=(v(y),\; 0 )
\end{equation}
with $v(y)$ a periodic function in $y$ can be deduced from 
(\ref{diffshe}) and reads
\begin{equation}
D^{E}_{11}=\sigma^2+\frac{1}{2 \pi} \int d k \; 
\frac{\mid \widehat{v}(k)\mid^2}{\sigma^2 k^2}
+ \tau \;\frac{1}{2 \pi} \int d k \; \mid \widehat{v}(k) 
\mid^2 + {\cal O}(\tau^2)\;\;\; ; \;\;\;
D^{E}_{12}=0 \;\;\; ; \;\;\; D^{E}_{22}=\sigma^2.
\label{esatta1}
\end{equation}
Applying the generalized formulas (\ref{chi0g}), (\ref{chi1g}) and
(\ref{diffgen}) to the shear flow we want to obtain the expressions
(\ref{esatta1}).

For the shear flow eq. (\ref{chi0g}) reads 

\begin{eqnarray}
(\partial_t + v\; \partial_1 -\sigma^2 \partial^2)w_1^{(0)}&=&-v 
\label{appb31}\\
(\partial_t + v\; \partial_1 -\sigma^2 \partial^2)w_2^{(0)}&=&0. 
\label{appb32}
\end{eqnarray} 
Inserting the stationary solution of eq. (\ref{appb32}),
$w_2^{(0)}=0$, into eq. (\ref{appb32}) and taking the Fourier
transform $\widehat{v}(k)$ of $v(y)$ we obtain:
    
\begin{equation}
w_1^{(0)}(y)=-\frac{1}{2 \pi} \int d k \; 
\frac{\widehat{v}(k)}{\sigma^2 \; k^2} \; e^{i k y}.
\label{w01}
\end{equation}
Thus eq. (\ref{chi1g}) becomes
\begin{eqnarray}
(\partial_t + v\; \partial_1 -\sigma^2 \partial^2)w_1^{(1)}&=&(p-q) \sigma^2 
\partial_2^2 v
\label{appb41}\\
 (\partial_t + v\; \partial_1 -\sigma^2 \partial^2)w_2^{(1)}&=&0. 
\label{appb42}
\end{eqnarray}
The  stationary solutions are   
\begin{equation}
w_1^{(1)}(y)=- (p-q) \; v(y) \;\;\; ; \;\;\;  w_2^{(1)}=0  \label{w11}.
\end{equation}
Using  eq. (\ref{diffgen}) the eddy diffusivity tensor is
\begin{eqnarray}
D^E_{12}&=&0\\
D^E_{22}&=&\sigma^2 \\
D^E_{11}&=&\sigma^2-\frac{1}{2}\langle v\; w_1^{(0)}\rangle -\frac{\tau}{2}
\left( q \sigma^2 \langle v \; \partial^2_2 w_1^{(0)}\rangle +
\langle v \; w_1^{(1)}\rangle\right)+{\cal O}(\tau^2)
\end{eqnarray}
and 
\begin{equation}
D^{E}_{11}=\sigma^2+\frac{1}{2 \pi}\int d k \; 
\frac{\mid \widehat{v}(k)\mid^2}{\sigma^2 k^2}
+ \tau \; (p-2q) \; \frac{1}{2 \pi}\int d k \; \mid \widehat{v}(k) \mid^2
+{\cal O}(\tau^2).
\label{gen1}
\end{equation}
A comparison between (\ref{gen1}) and (\ref{esatta1}) shows that the
generalized equations lead to the exact expression of the eddy
diffusivity tensor for a stationary shear flow if $$ p=2q+1.$$ The
same condition is found if we consider a time-dependent shear flow.


\newpage
\narrowtext

\end{document}